\documentstyle[amsmath,amssymb, 
aps,epsfig,twocolumn]{revtex}

\begin{document}
 \tolerance 50000 \def\d{\dagger} \draft

\twocolumn[\hsize\textwidth\columnwidth\hsize%
\csname@twocolumnfalse\endcsname

\title{A Family of Grover's Quantum Searching Algorithms }

\author{A. Galindo and M.A. Mart\'{\i}n-Delgado } \address{ 
Departamento de F\'{\i}sica Te\'orica I, Universidad 
Complutense, 28040-Madrid, Spain.  }

\maketitle

\begin{abstract}
We introduce the concepts of Grover operators and Grover kernels
to systematically analyse Grover's searching algorithms. Then,
we investigate a one-parameter family of quantum searching 
algorithms of Grover's type and we show that the standard 
Grover's algorithm is a distinguished member of this family.  
We show that all the algorithms of this class solve the 
searching problem with an efficiency of order $O(\sqrt{N})$, 
with a coefficient which is class-dependent.  The analysis 
of this dependence is a test of the stability and robustness 
of the algorithms. We show the stability of this constructions
under perturbations of the initial conditions and extend them
upon a very general class of Grover operators.
\begin{center}
\parbox{14cm}{}

\end{center}
\end{abstract}

\vspace{-0.8 true cm}

\pacs{ \hspace{2.5cm} PACS number: 03.67.Lx, 03.67.-a }

\vskip2pc] \narrowtext

\vspace{-1 true cm}

\section{Introduction}

The problem of searching an element in a list of $N$ 
unsorted elements when this number becomes very large is 
known to be one of the basic problems in Computational 
Science.  Classically, one may devise many strategies to 
perform that search, but if the elements in the list are 
distributed with equal probability, then we shall need to 
make $O(N)$ trials in order to have a high confidence of 
finding the desired element, also called marked element.  
The formulation of quantum computation as a well-stablished 
theoretical discipline for storing and processing 
information \cite{QC} has opened the possibility of designing new 
searching algorithms with no classical analogue.
The familiar Grover quantum searching algorithm takes 
advantage of the quantum mechanical properties to perform 
the searching problem with an efficiency of order 
$O(\sqrt{N})$ \cite{grover1}, \cite{grover2}.

In classical computation there exist design techniques which 
provide general directions for algorithmic problem solving.  
In quantum computation however, the list of quantum 
algorithms is very short.  It seems we are lacking the basic 
principles underlying the quantum algorithmic design.  Under 
these circumstances it is a good choice to put to the test 
the currently known quantum algorithms.  In this work our 
purpose is to follow this goal with Grover's quantum 
searching algorithm by trying to understand the relevant 
pieces of this algorithm and wondering to what extent they 
allow for generalization \cite{boyer}, \cite{zalka}, 
\cite{biham1}, \cite{biham2}, \cite{jozsa}, \cite{mda}.

Let us state the searching problem in terms of a list ${\cal 
L}[0,1,\ldots, N-1]$ with a number $N$ of unsorted elements.  
We shall denote by $x_0$ the marked element in ${\cal L}$ 
that we are searching for.  The quantum mechanical solution 
of this searching problem goes through the preparation of a 
quantum register in a quantum computer to store the $N$ 
items of our list.  This is how quantum parallelism is 
realized.  Thus, let us assume that our quantum registers 
are made of $n$ qubits so that the total elements we have 
are $N=2^n$.  Let us denote by $|x\rangle$, 
$x=0,1,\ldots,N-1$ the ket states of the computational basis 
which are orthonormalized.  Any state $|\Psi\rangle$ of the 
quantum register is a linear superposition of the 
computational states.  In the beginning of the algorithm the 
quantum register is initialized to a given quantum state 
$|\Psi\rangle = |x_{\rm in}\rangle$.

The second component of the algorithm is to design a quantum 
operation which will be repeatedly applied to $|x_{\rm 
in}\rangle$ in order to find the marked element.  This 
strategy is similar to the classical counterpart algorithm.  
The difference is the fact that the quantum operation is 
realized in terms of an unitary operator which implements 
the reversible quantum computation.  It is this quantum 
operation what has been so neatly designed by Grover \cite{grover1}.  With 
Grover's choice we may say that the quantum evolution is 
such that the constructive interference of quantum 
amplitudes is directed towards the marked state one looks 
for.

\section{Grover Operators}

In order to set up our analysis we shall need to introduce 
some definitions.

\noindent {\em Definition 1} A Grover operator $G$ is any 
unitary operator with at most two different eigenvalues; 
i.e., $G$ a linear superposition of two orthogonal 
projectors $P$ and $Q$:
\begin{equation}
G = \alpha P + \beta Q, \; \; P^2 = P, \; Q^2 = Q, \; \; P + 
Q = 1
\label{1}
\end{equation}
where $\alpha, \beta \in {\bf C}$ are complex numbers of 
unit norm.

\noindent {\em Definition 2} A Grover kernel $K$ is the 
product of two Grover operators:
\begin{equation}
K = G_2 G_1
\label{2}
\end{equation}
Some elementary properties follow immediately from these 
definitions.

\noindent {\em Property 1} Any Grover kernel $K$ is a 
unitary operator, and therefore, it can be used to implement 
the unitary evolution in a quantum computer.

\noindent {\em Property 2} Let the Grover operators $G_1, 
G_2$ be chosen such that
\begin{equation}
G_1 = \alpha P_{x_0} + \beta Q_{x_0}, \quad P_{x_0} = 
|x_0\rangle \langle x_0|, \; P_{x_0} + Q_{x_0} = 1
\label{3}
\end{equation}
\begin{equation}
G_2 = \gamma \bar{P} + \delta \bar{Q}, \quad \bar{P} + 
\bar{Q} = 1
\label{4}
\end{equation}
with $\bar{P}$ given by the rank 1 matrix
\begin{equation}
\bar{P} = {1\over N} \left(
\begin{array}{ccc}
1 & \ldots & 1 \\
\vdots & & \vdots \\
1 & \ldots & 1
\end{array}
\right)
\label{5}
\end{equation}

\noindent This is clearly a projector $\bar{P} = |k_0\rangle 
\langle k_0|$ on the subspace spanned by the state 
$|k_0\rangle = {1\over \sqrt{N}} (1,\ldots ,1)^{\rm t}$, 
where the superscript denotes the transpose.  Then, if we 
take the following set of parameters,
\begin{equation}
\alpha = -1, \; \beta = 1, \; \gamma = -1, \; \delta = 1
\label{6}
\end{equation}

\noindent the Grover kernel (\ref{2}) reproduces the 
original Grover's choice.  This property follows inmediately 
by construction.  In fact, we have in this case $G_1 = 1 - 2 
P_{x_0}=:G_{x_0}$ whilst the operator $G_2 = 1 - 2 \bar{P}$ 
coincides with the diffusion operator introduced by Grover 
to implement the inversion about the average \cite{grover1}.

One can also show the following 
property which provides a geometrical meaning for the Grover kernels.

\noindent {\em Property 3} Let ${\cal K}_{\rm G}$ denote
the set of all the Grover kernels for fixed
$\{|x_0\rangle,|k_0\rangle\}$.  Then ${\cal K}_{\rm G}$ can
be viewed as a 3D-subset of the group U(2) which is of the form
$S^1\times {\cal K}_{\rm G}^\prime$, where ${\cal K}_{\rm
G}^\prime$ is a 2D-submanifold of SU(2) (Fig.1).

\begin{figure}
\begin{center}
\epsfig{file=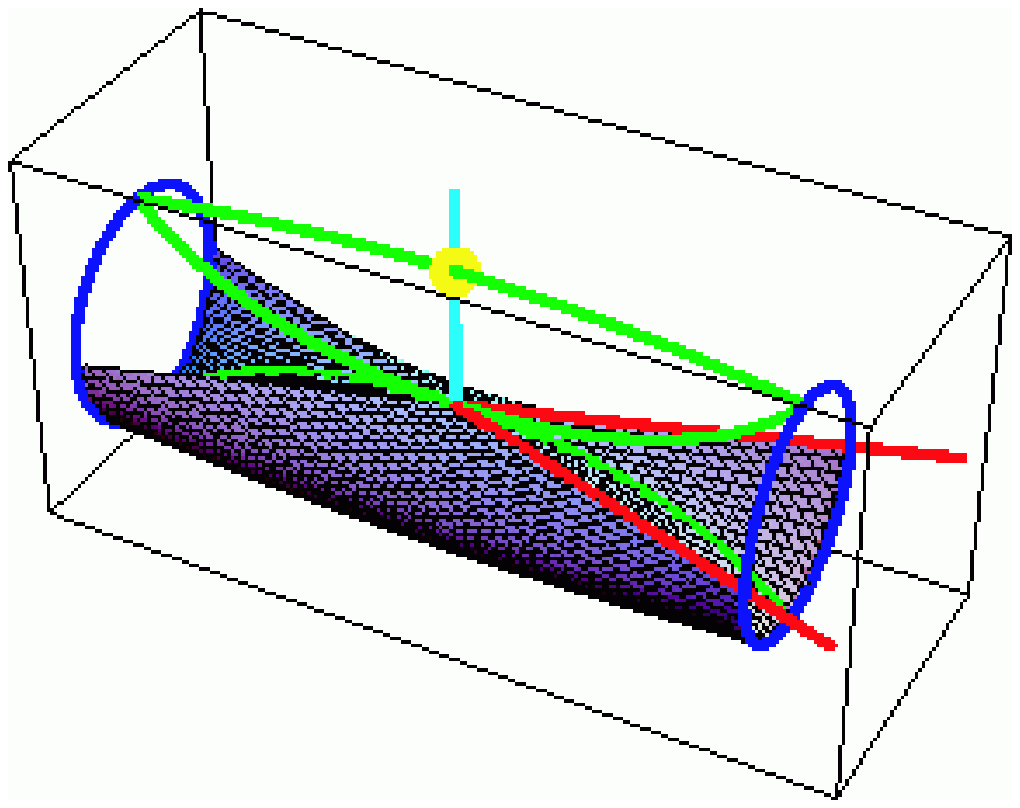,width=0.45\linewidth}\hfill
\epsfig{file=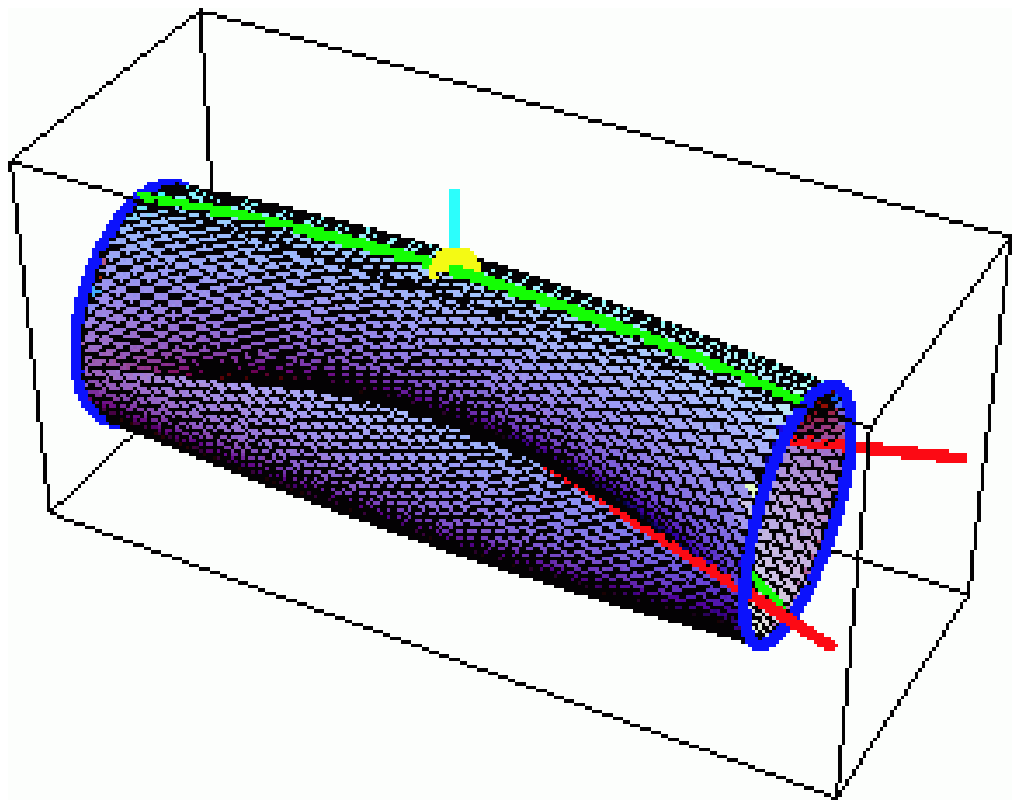,width=0.45\linewidth}\\
\vskip 0.25cm
\narrowtext \caption[]{Left: half of the surface ${\cal
K}_{\rm G}^\prime$ for $N=10$.  Right: the whole surface ${\cal
K}_{\rm G}^\prime$ for $N=10$. In red, the rotation axes
corresponding to ${\rm i}G_{1,2}\in SU(2)$. In pale blue, the normal
to their plane.  The yellow dot signals the original Grover
kernel.  The blue curves are the border $\alpha=\pi,3\pi$, 
where $\alpha$ is the rotation angle. In green, the locus of 
kernels $K$ with factors ${\rm i}G_{1,2}$ with the same 
rotation angles mod $2\pi$.}
\label{fig1}
\end{center}
\end{figure}


The content of this property is illustrated in Fig.1.
It follows from the fact that the two  parameters of a  Grover kernel
in Definition 2 with the fixing (\ref{6}) can be used to parametrize 
a subset of the  unitary group $U(2)$ of complex $2\times 2$ matrices.
This 3D-subset has a factorized form $S^1\times {\cal K}_{\rm G}^\prime$
where $S^1$ is the unit circle (the group U(1) and 
we call   ${\cal K}_{\rm G}^\prime$ a certain 2D-submanifold of the
group of special unitary matrices SU(2) whose construction we
explain in the following and we plot in Fig.1.
This figure is constructed by parametrizing the two elements
$\pm {\rm i}G_{1,2}$ of SU(2) as follows, 
$\pm {\rm i}G_{1,2}={\rm e}^{{\rm i} \alpha_{1,2} {\bf n}_{1,2}\cdot 
\mathbf{\sigma}}$,
where $\mathbf{\sigma}$ are the Pauli matrices and ${\bf n}_{1,2}$ are 
unit vectors which are kept fixed.
Likewise, we parametrize the corresponding Grover kernel (\ref{2})
as $K={\rm e}^{{\rm i} \alpha {\bf n}\cdot \mathbf{\sigma}}$.
Then, upon varying the parameters $\alpha_{1,2}$ we obtain
the surface depicted in Fig.1 


\noindent Let us point out the following interpretation of 
the Grover operators.  Let us think of the computational 
basis $\{ |x\rangle \}$ as a coordinate basis in Quantum 
Mechanics and introduce the quantum discrete Fourier 
transform in the standard fashion, $|\hat x\rangle:=U_{\rm 
DFT} |x \rangle = {1\over \sqrt{N}} \sum_{y=0}^{N-1} {\rm 
e}^{2 \pi {\rm i} x\cdot y/N} |y\rangle$.  The transformed 
basis $\{ |\hat x\rangle \}$ can then be seen as the dual momentum 
basis.  Then, it is easy to see that in such a basis the 
projector operator $\bar{P}$ takes the following form:
\begin{equation}
U_{\rm DFT}^{-1} \bar{P} U_{\rm DFT} = |0\rangle \langle 0| 
=:P_{0}
\label{7}
\end{equation}

\noindent This means that the Grover operator $G_2$ takes 
the same matricial form in the momentum basis as the Grover 
operator $G_1$ in the coordinate basis.  They are somehow 
dual of each other.  The original Grover kernel takes then the 
form
\begin{equation}
K = U_{\rm DFT} G_{x=0} U_{\rm DFT}^{-1} G_{x_0}
\label{8}
\end{equation}
\noindent which shows that a Grover kernel has a part local 
in coordinate space and anoter part which is local in 
momentum space. 

\noindent This ``momentum" interpretation of the search algorithm
stems from a quantum mechanical analogy between the computational
basis and its Fourier transformed states which enter the definition
of the Grover operators.
We would like to point out that similar analogies have been used
in connection with alternative formulations of the quantum searching
algorithm, namely, the analog analogue of a digital quantum computation
with Grover's algorithm \cite{farhi-gutmann} which is based in a Hamiltonian
formulation.

\section{The Searching Algorithms}

\subsection{The Basic Formalism}

Next, the third part of the algorithms corresponds to 
applying the Grover kernel $K$ to the initial state $|x_{\rm 
in}\rangle$ a number of times $m$ seeking a final state 
$|x_{\rm f}\rangle$ given by

\begin{equation}
|x_{\rm f}(m)\rangle = K^m |x_{\rm in}\rangle
\label{9}
\end{equation}

\noindent such that the probability ${\cal P}(x_0)$ of 
finding the marked state is above a given threshold value.  
We shall take this value to be $1/2$, meaning that we choose 
a probability of success of $50 \%$ or larger.  Thus, we are 
seeking under which circumstances the following condition

\begin{equation}
{\cal P}(x_0) = |\langle x_0 | K^m |x_{\rm in}\rangle|^2 > 
1/2
\label{10}
\end{equation}

\noindent holds true.

The analysis of this probability gets simplified if we 
realize that the evolution associated to the searching 
problem can be mapped onto a reduced 2D-space spanned by the 
vectors

\begin{equation}
\{|x_0\rangle, |x_{\perp}\rangle := {1\over {\sqrt{N-1}}} 
\sum_{x\neq x_0} |x\rangle\}
\label{11}
\end{equation}

\noindent Then we can easily compute the projections of the 
Grover operators $G_1,G_2$ in the reduced basis with the 
result
\begin{equation}
G_1 = \left(
\begin{array}{cc}
\alpha & 0 \\
0 & \beta
\end{array}
\right)
\label{12}
\end{equation}
\begin{equation}
G_2 = \left(
\begin{array}{cc}
\delta & 0 \\
0 & \gamma
\end{array}
\right) + (\gamma - \delta) \left(
\begin{array}{cc}
{1\over N} & {\sqrt{N-1}\over N} \\
{\sqrt{N-1}\over N} & {-1\over N}
\end{array}
\right)
\label{13}
\end{equation}

\noindent From now on, we shall fix two of the phase 
parameters using the freedom we have to define each Grover 
factor in (\ref{2}) up to an overall phase.  Then we decide 
to fix them as follows:

\begin{equation}
\alpha = \gamma = -1
\label{14}
\end{equation}

\noindent With this choice, the Grover kernel (\ref{2}) 
takes the following form in this basis:

\begin{equation}
K = {1\over N} \left(
\begin{array}{cc}
1 + \delta (1-N) & -\beta (1+\delta) \sqrt{N-1} \\
(1+\delta) \sqrt{N-1} & \beta (1+\delta-N)
\end{array}
\right)
\label{15}
\end{equation}

We shall fix the initial conditions using the same initial 
state $|x_{\rm in}\rangle$ as in the original Grover's 
algorithm \cite{grover1}, i.e., we choose the uniform state corresponding 
to zero momentum and find its components in the reduced 
basis to be
\begin{equation}
|x_{\rm in}\rangle = {1\over {\sqrt{N}}} |x_0\rangle + 
\sqrt{{N - 1\over N}} |x_{\perp}\rangle
\label{16}
\end{equation}

In order to compute the probability amplitude in (\ref{10}), 
we introduce the spectral decomposition of the Grover kernel 
$K$ in terms of its eigenvectors $\{ |\kappa_1\rangle, 
|\kappa_2\rangle \}$, with eigenvalues ${\rm e}^{{\rm 
i}\omega_1}, {\rm e}^{{\rm i}\omega_2}$.  Thus we have
\begin{equation}
    \begin{split}
&{\cal A}(x_0) := \langle x_0 | K^m |x_{\rm in}\rangle = \\
&{1\over \sqrt{N}} \sum_{j=1}^2 \left\{|\langle 
x_0|\kappa_j\rangle|^2 + {\sqrt{N-1}} \langle 
x_0|\kappa_j\rangle \langle 
\kappa_j|x_{\perp}\rangle\right\} {\rm e}^{{\rm i}m\omega_j}
\label{17}
    \end{split}
\end{equation}

\noindent This in turn can be casted into the following 
closed form:
\begin{equation}
    \begin{split}
&\langle x_0 | K^m |x_{\rm in}\rangle = \\
&{\rm e}^{{\rm i}m\omega_1} \left( {1\over \sqrt{N}} + ({\rm 
e}^{{\rm i}m \Delta \omega} - 1) \langle x_0|\kappa_2\rangle 
\langle \kappa_2|x_{\rm in}\rangle \right)
\label{18}
    \end{split}
\end{equation}

\noindent with $\Delta \omega = \omega_2 - \omega_1$.

In terms of the matrix invariants
\begin{equation}
{\rm Det} K = \beta \delta, \quad {\rm Tr} K = - (\beta + 
\delta) + (1+\beta)(1+\delta) {1\over N}
\label{19}
\end{equation}
the eigenvalues $\zeta_{1,2} = {\rm e}^{{\rm 
i}\omega_{1,2}}$ are given by
\begin{equation}
\zeta_{1,2} = ({\rm Tr} K)/2 \mp \sqrt{-{\rm Det} K + ({\rm 
Tr} K/2)^2}
\label{20}
\end{equation}
The corresponding unnormalized eigenvectors are
\begin{equation}
|\kappa_{1,2}\rangle \propto \left(
\begin{array}{c}
{A \mp \sqrt{-4({\rm Det} K) N^2 + A^2}\over 
2(1+\delta)\sqrt{N-1}} \\
1
\end{array}
\right)
\label{21}
\end{equation}

\noindent with
\begin{equation}
A := (\beta - \delta) N + (1-\beta)(1+\delta)
\label{22}
\end{equation}

\noindent Although we could work out all the expressions for 
a generic value $N$ of elements in the list, we shall 
restrict our analysis to the case of a large number of 
elements, $N \rightarrow \infty$, and we shall leave for a 
numerical simulation the effect of arbitrary $N$.  Thus, in 
this asymptotic limit we need to know the behaviour for 
$N\gg 1$ of the eigenvector $|\kappa_2\rangle$ which turns 
out to be

\begin{equation}
|\kappa_{2}\rangle \propto \left(
\begin{array}{c}
{\beta - \delta \over 1 + \delta} \sqrt{N} + 
O({1\over\sqrt{N}}) \\
1
\end{array}
\right)
\label{23}
\end{equation}

\noindent Thus, for generic values of $\beta, \delta$ we 
observe that the first component of the eigenvector 
dominates over the second one meaning that asymtoptically 
$|\kappa_2\rangle \sim |x_0\rangle$ and then $\langle 
x_0|\kappa_2\rangle \langle \kappa_2|x_{\rm in}\rangle = 
O({1\over \sqrt{N}})$.  This implies that the probability of 
success in (\ref{18}) will never reach the threshold value 
(\ref{10}).  Then we are forced to tune the values of the 
two parameters in order to have a well-defined and 
nontrivial algorithm and we demand
\begin{equation}
\beta = \delta\neq -1
\label{24}
\end{equation}

Now the asymptotic behaviour of the eigenvector changes and 
is given by a balanced superposition of marked and unmarked 
states, as follows

\begin{equation}
|\kappa_{2}\rangle \sim {1\over\sqrt{2}} \left(
\begin{array}{c}
{\rm i}\delta^{1/2} \\
1
\end{array}
\right)
\label{25}
\end{equation}

\noindent This is normalized and we see that none of the 
components dominates.  When we insert this expression into 
(\ref{18}) we find

\begin{equation}
|\langle x_0 | K^m |x_{\rm in}\rangle| \sim {|\delta |\over 
2} |{\rm e}^{{\rm i}m \Delta \omega} - 1| \sim \left|\sin 
({m \Delta \omega \over 2})\right|
\label{26}
\end{equation}

\noindent This expression means that we have succeded in 
finding a class of algorithms which are apropriate for 
solving the quantum searching problem.  Now we need to find 
out how efficient they are.  To do this let us denote by $M$ 
the values of the time step $m$ at which the probability 
becomes maximum; then

\begin{equation}
M = \lfloor\left|\pi /\Delta \omega\right|\rfloor
\label{27}
\end{equation}

\begin{figure}
\begin{center}
\epsfig{file=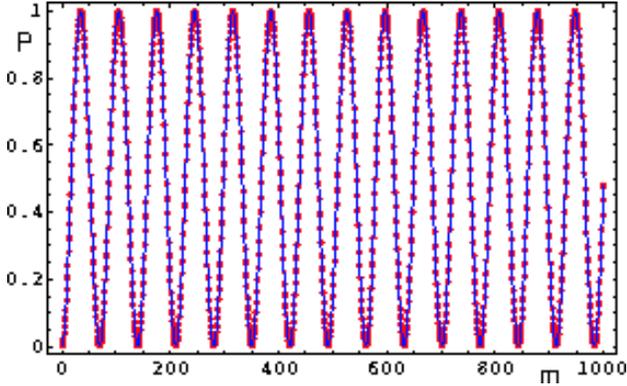,width=0.95\linewidth} \vskip 0.25cm
\narrowtext \caption[]{Probability of success ${\cal P}$ as
a function of the time step for $N=1000$ and
$\beta=\delta={\rm e}^{{\rm i}\pi/2}$.}
\label{fig2}
\end{center}
\end{figure}


\noindent As it happens, we are interested in the asymptotic 
behaviour of this optimal periods of time $M$.  From the 
equation (\ref{20}) we find the following behaviour as $N 
\rightarrow \infty$:

\begin{equation}
\Delta \omega \sim {4 \over \sqrt{N}}{\rm Re}\sqrt{\delta}
\label{28}
\end{equation}

\noindent Thus, if we parametrize $\delta = {\rm e}^{{\rm 
i}\phi}$, then we finally obtain the expression,

\begin{equation}
M \sim \left\lfloor {\pi \over 4 \cos {\phi \over 
2}}\sqrt{N}\right\rfloor
\label{29}
\end{equation}

Therefore, we conclude that the Grover algorithm of the 
class parametrized by $\phi$ is a well-defined quantum 
searching algorithm with an efficiency of order 
$O(\sqrt{N})$ and with a subdominant behaviour which depends 
on each element of the family.  Within this class, the 
original Grover algorithm is a distinguished element for 
which the coefficient in (\ref{29}) achieves its optimal 
value at $\phi = 0$.  Moreover, the worst value occurs for 
$\phi \to\pi$; in this limit $M$ is not well-defined, and it 
corresponds to trivial case where the Grover kernel is just 
the identity operator, $K=1$.

\noindent The expression (\ref{29}) for $M$ can also be 
given another meaning regarding the stability of the 
Grover's case $\phi = 0$.  It is plain that under a small 
perturbation $\delta\phi$ around this value, its optimal 
nature is not spoiled in first order for we find a behaviour 
which is quadratic in the perturbation, namely, $M \sim {\pi 
\over 4 } (1 + 0.125(\delta\phi)^2)\sqrt{N}$.
This  stability considered here 
is with respect to perturbations in eigenvalues (or
eigenvectors) in the reduced 2-dimensional subspace
specified by the quantum searching problem (\ref{11}). 
We also require these type of perturbations to hold
in all iterations.

\noindent Howewer, if we happen to choose a Grover kernel 
with a $\phi$ far from $0$ we may end up with a searching 
algorithm for which the leading behaviour order 
$O(\sqrt{N})$ is masqueraded by the big value of the 
coefficient and the time to achive a succedding probability 
becomes very large.  For instance, we may have a Grover 
kernel with a behaviour $M \sim 10^3\sqrt{N}$ and for a 
value of $N=10^6$ it would turn out as efficient as a 
classical algorithm of order $O(N) = 10^6$.  Thus, the limit 
$\phi \rightarrow \pi$ behaves as a sort of classical limit 
where the quantum properties disappear.

\begin{figure}
\begin{center}
\epsfig{file=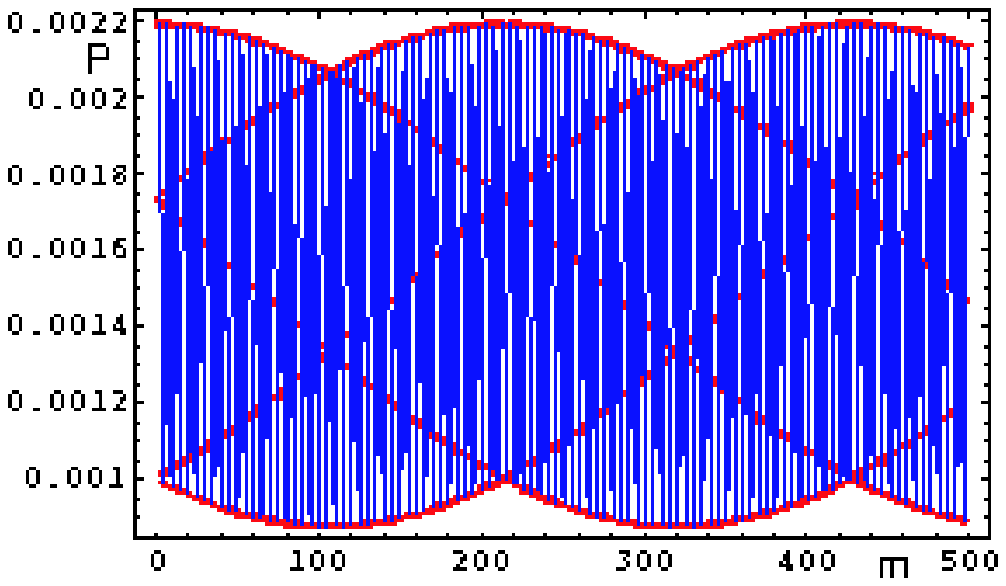,width=0.95\linewidth} \\
\epsfig{file=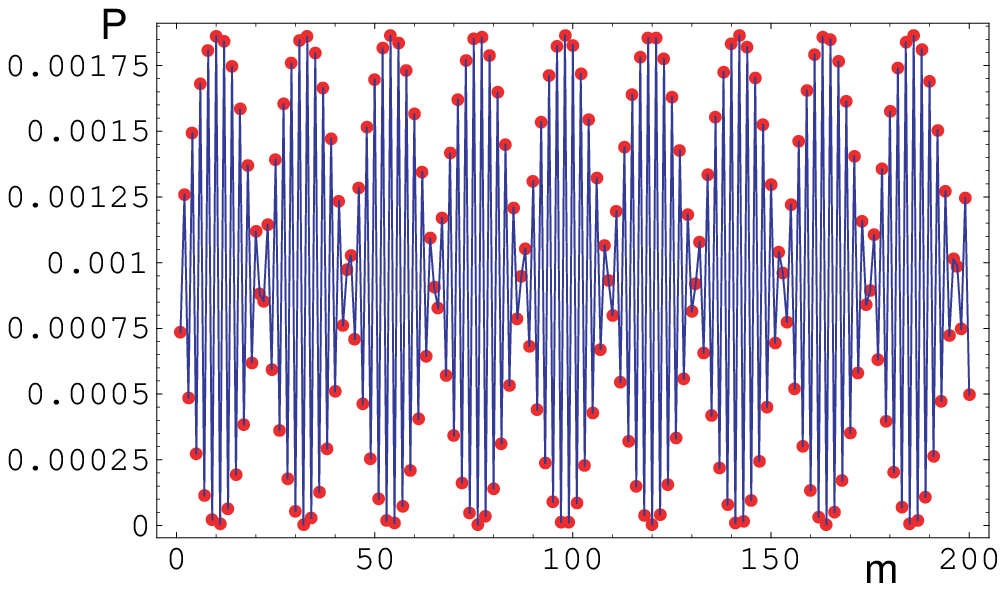,width=0.95\linewidth} 
\vskip 
0.25cm
\narrowtext \caption[]{Probability of 
success ${\cal P}$ as a function of the time step for 
$N=1000$ and $\beta={\rm i}, \delta={\rm
i}{\rm e}^{{\rm i}5/4}$ (up), $\beta={\rm i}, \delta={\rm i}{\rm
e}^{{\rm i}3}$ (down). }
\label{fig3}
\end{center}
\end{figure}

\noindent
In Fig. 2 we have plotted the probability of success as a 
function of the time step $m$ for a list of $N=1000$ 
elements and a choice of parameters $\beta=\delta={\rm 
e}^{{\rm i}\pi/2}$ satisfying condition (24).  We observe 
how the algorithm is fully efficient in achieving the 
maximum probability possible.  Despite $\phi=\pi/2$ is not 
close to Grover's optimal value of $0$, we find an excellent 
behaviour.  The main difference with the optimal case is 
that here the number of maxima is $14$ while for Grover's it 
is $20$, as implied by (\ref{29}).
This looks like a pattern of fully constructive 
interference.

In Fig. 3 we have plotted the same function but
with a choice of parameters $\beta={\rm i}, \delta={\rm
i}{\rm e}^{{\rm i}5/4}$ ($\beta={\rm i}, \delta={\rm i}{\rm
e}^{{\rm i}3}$) violating condition (24).  We observe how
the algorithm becomes inefficient and the maximum it takes
is lesser than $0.0021923$ ($0.001864$) for any time step. 
This looks like a pattern of partially constructive
interference.


We find this behaviour as reminiscent of a quantum phase 
transition where the transition is driven by quantum 
fluctuations instead of standard thermal fluctuations.  In 
this type of transition each quantum phase is characterized 
by a ground state which is different in each phase.  It is 
the variation of a coupling constant in the Hamiltonian of 
the quantum many-body problem which controls the occurrence 
of one quantum phase or another in the same manner as the 
temperature does the job in thermal transitions.  In our 
case we may consider the two different asymptotic behaviours 
of the eigenvector $|\kappa_2\rangle$ as playing the role of 
two ground states.  Following this analogy, we may see our 
family of algorithms parametrized by a torus ${\cal T} = S^1 
\times S^1$ where the parameters $\beta$ and $\delta$ take 
their values and the difference $g := \beta - \delta$ is a 
sort of coupling constant which governs in which of the two 
phases we are.  When $g\neq 0$ we fall into a sort of 
disordered phase where the efficiency of this class of 
Grover's algorithms is spoiled.  However, when $g=0$ we are 
located precisely at one equal superposition of the pricipal 
cycles of the torus which defines a one-parameter family of 
efficient algorithms.


\subsection{The Influence of Initial Conditions}

Next we shall address the issue of to what extent this 
one-parameter family of algorithms depends on the choice of 
initial conditions for the initial state $|x_{\rm 
in}\rangle$.  We would like to check that the stable 
behaviour we have found is not disturbed under perturbations 
of initial conditions.

\noindent Let us consider a more general initial state 
$|x_{\rm in}\rangle$ which is not the precise one used in 
the original Grover's algorithm \cite{grover1} but instead it is chosen as

\begin{equation}
|x_{\rm in}\rangle = {a \over {\sqrt{N}}} |x_0\rangle + b 
\sqrt{{N - 1\over N}} |x_{\perp}\rangle
\label{30}
\end{equation}

\noindent where $a$ and $b$ are chosen to satisfy a 
normalization condition.  Then, it is possible to go over 
the previous analysis and find that the probability 
amplitude is now given by

\begin{equation}
    \begin{split}
&\langle x_0 | K^m |x_{\rm in}\rangle = \\
&{\rm e}^{{\rm i}m\omega_1} \left( {a\over \sqrt{N}} + ({\rm 
e}^{{\rm i}m \Delta \omega} - 1) \langle x_0|\kappa_2\rangle 
\langle \kappa_2|x_{\rm in}\rangle \right)
\label{31}
    \end{split}
\end{equation}

\noindent where now $|x_{\rm in}\rangle$ is the new initial 
state (\ref{30}).  We have to distinguish two cases: i) 
The coefficient $a$ of the marked state is order 1 and ii) 
it is order bigger than 1, say of order 
$O(\sqrt{N})$.  In the latter case ii), it means that the 
initial state is so peaked around the marked state that we 
do not even need to resort to a searching algorithm, but 
instead measure directly on the initial state to find 
sucessfully the marked state.  Thus, we shall restrict to 
case i) in the following.  Now the key point is to realize 
that all the previous asymptotic analysis is dominated by 
the behaviour of the eigenvector $|\kappa_2 \rangle$ given 
by expression (\ref{23}) which is something intrinsic to the 
Grover kernel and independent of the initial conditions.  
Thus, if condition (\ref{24}) is not satisfied, then as we 
are in case i) the first term in the RHS of (\ref{31}) is 
not relevant and we are led again to the conclusion that the 
algorithm is not efficient.  On the contrary, if condition 
(\ref{24}) is satisfied the same mechanism based on 
(\ref{25}) operates again and the algorithm has a 
probability of success measured by

\begin{equation}
|\langle x_0 | K^m |x_{\rm in}\rangle| \sim |b|\sin ({m 
\Delta \omega \over 2})
\label{32}
\end{equation}

\noindent with $\Delta \omega$ also given by (28).  Then we 
may conclude that the class of algorithms is stable under 
perturbations of the initial conditions.

\subsection{Extended Formalism}

Finally, we would like to check how general is this 
construction in terms of projection operators of the type 
used in (5) for $\bar{P}$.  To this end let us recall that 
$\bar{P}$ can be interpreted as the projector $|\hat 0\rangle 
\langle \hat 0|$.  Thus a natural generalization is to 
consider a projector on a different momentum state, say 
$|\hat y_0\rangle$, with $y_0\neq 0$.  The matrix elements 
of this projection operator in the coordinate basis are

\begin{equation}
(\bar{P})_{x,x'} = {1\over N} {\rm e}^{2\pi{\rm i} (x' - 
x)\cdot y_0/N}, \quad x,x' = 0,1,\ldots, N-1
\label{33}
\end{equation}

We can go even further and consider a general form for the 
states $|x_0\rangle, |x_{\perp}\rangle, |k_0\rangle$ as 
follows,

\begin{equation}
\begin{split}
&|x_0\rangle = (1,0,\ldots, 0)^{\rm t} \\  
&|x_{\perp}\rangle = {1\over \sqrt{1 - 
\alpha_1^2}} (0,\alpha_2,\ldots, 
\alpha_N)^{\rm t} \\ 
&|k_0\rangle = 
(\alpha_1,\alpha_2,\ldots, \alpha_N)^{\rm t}
\label{34}
\end{split}
\end{equation}

\noindent where there is no loss of generality by chosing 
$|x_0\rangle$ in this way; $\alpha_1,\ldots,\alpha_N$ is a 
given and normalized set of arbitrary complex amplitudes, 
with $\alpha_1>0$. We will assume that $||\alpha ||^2 > 
\alpha_1^2$.  

\noindent The projector $\bar{P}$ is chosen to be

\begin{equation}
\bar{P} = |k_0\rangle \langle k_0|
 \label{35}
\end{equation}
and it admits (\ref{33}) as a particular case.

Now in the reduced 2D-basis spanned by $\{ 
|x_0\rangle, |x_{\perp}\rangle\}$ the Grover kernel has
the following expression:

\begin{equation}
K = 
\left(
\begin{array}{cc}
-\delta+ \Delta \alpha_1^2 & -\beta \Delta 
\alpha_1 \sqrt{1-\alpha_1^2} \\
\Delta \alpha_1\sqrt{1-\alpha_1^2} & 
\beta (\Delta \alpha_1^2 - 1)
\end{array}
\right)
\label{36}
\end{equation}

\noindent with $\Delta :=1 + \delta$.  Thus all the 
dynamics depends on the relative strength of the real amplitude 
$\alpha_1$ with respect to the rest of the amplitudes.  If 
we set $\alpha_i = 1/\sqrt{N}$ $\forall i$, then we recover the same 
expression as in (\ref{15}).  Moreover, the initial 
condition is taken as

\begin{equation}
|x_{\rm in}\rangle = \alpha_1 |x_0\rangle + 
\sqrt{1- \alpha_1^2} |x_{\perp}\rangle
\label{37}
\end{equation}

\noindent In order to perform our analysis, we shall assume
that the unknown amplitude $\alpha_1$ behaves generically as
$\alpha_1\sim 1/\sqrt{N}$, and consequently $\sqrt{1 -
\alpha_1^2} \sim \sqrt{1-1/N}$.  Under these circumstances,
we find the following asymptotic behaviour for the
eigenvector $|\kappa_2\rangle$ of the Grover kernel: if
$\beta \neq \delta$ and $\delta\neq -1$,

\begin{equation}
|\kappa_{2}\rangle \propto \left(
\begin{array}{c}
{\beta - \delta \over 1 + \delta} {1\over 
\alpha_1} \\
1
\end{array}
\right)
\label{38}
\end{equation}

\noindent and if $\beta = \delta$,
\begin{equation}
|\kappa_{2}\rangle \sim {1\over \sqrt{2}} \left(
\begin{array}{c}
{\rm i}\delta^{1/2} \\ 
1
\end{array}
\right)
\label{39}
\end{equation}

\noindent This latter case is again the only favorable to 
obtain an efficient algorithm and the behaviour of the time 
$M$ for achieving maximum probability of success takes the 
following form

\begin{equation}
M \sim {\pi\alpha_1^{-1} \over 4 \cos {\phi \over 2}}
\label{40}
\end{equation}

We conclude then that our construction of quantum searching 
algorithms of Grover's type are general enough under 
different choices of Grover operators $G_1,G_2$ and that the 
analyisis performed with the simplest choice of these 
operators captures the essential properties of the class of 
algorithms we have presented.

\section{Conclusions}

We have introduced the notion of Grover operators and Grover kernels
which lead to  a systematic study of Grover's quantum searching
algorithms. These notions facilitates the generalization of Grover's 
algorithms in several direcctions. We have characterized the basic
features of these algorithms in terms of these operarators whose main
properties we have established in Sect. II.
Using these operators we have investigated a family of Grover kernels
whose qualities  as efficient algorithms 
depend on the range of parameters entering  the construction of their
associated Grover operators. When the algorithms are efficient, they
also perform the searching task with order $O(\sqrt{N})$, and the original
Grover's choice gives the  optimum value 
in the one-paramater family of algorithms.
Moreover, we have extended this study to incorporate  initial
conditions different than the standard uniform initial states 
and we have checked
that letting aside exceptional cases, the basic algorithms of Sect. III
maintain their  efficiency.
Finally, we have addressed  also the issue  of considering quite general 
Grover operators 
and found that the basic efficiency properties of the simplest choice's
for Grover's algorithm remain unchanged.

{\bf Acknowledgements} We would like to thank J.I. Cirac and L.K. Grover
for carefully reading the manuscript and suggesting new references.
We are partially supported by the 
CICYT project AEN97-1693 (A.G.) and by the 
DGES spanish grant PB97-1190 (M.A.M.-D.).

\vspace{-0.5 true cm}

\end{document}